%% file: puncturing.tex
\documentclass[journal]{IEEEtran}

\newtheorem{theorem}{Theorem}
\newtheorem{definition}{Definition}

\usepackage{algorithmic}
\usepackage{cite}
\usepackage{graphicx}
\usepackage{amsfonts}
\usepackage{mathtools}

\title{Untainted Puncturing for Irregular Low-Density Parity-Check Codes}

\author{
  \IEEEauthorblockN{David Elkouss, Jesus Martinez-Mateo, and Vicente Martin}
\thanks{
The authors are with the Facultad de Inform\'{a}tica, Universidad Polit\'{e}cnica de Madrid, Spain (e-mail: \{delkouss, jmartinez, vicente\}@fi.upm.es).}
}

\begin{document}
\maketitle

\input{abstract}

\IEEEpeerreviewmaketitle

\input{introduction}

\input{notation}

\input{properties}

\input{algorithm}

\input{results}

\input{conclusions}

\section*{Acknowledgment}

This work has been partially supported by the project Quantum Information Technologies in Madrid 
(QUITEMAD), Project P2009/ESP-1594, \textit{Comunidad Aut\'onoma de Madrid}.

The authors would like to thank the assistance and computation resources provided by \textit{Centro de Supercomputaci\'on y Visualizaci\'on de Madrid}\footnote{http://www.cesvima.upm.es} (CeSViMa).

\bibliographystyle{IEEEtran}
\bibliography{puncturing}

\vfill

\end{document}

%% file: abstract.tex
\begin{abstract}
Puncturing is a well-known coding technique widely used for constructing rate-compatible codes. In this paper, we consider the problem of puncturing low-density parity-check codes and propose a new algorithm for intentional puncturing. The algorithm is based on the puncturing of \textit{untainted} symbols, i.e. nodes with no punctured symbols within their neighboring set. It is shown that the algorithm proposed here performs better than previous proposals for a range of coding rates and short proportions of punctured symbols.
\end{abstract}

\begin{IEEEkeywords}

low-density parity-check codes, intentional puncturing, short-length codes.

\end{IEEEkeywords}

%% file: introduction.tex
\section{Introduction}

Low-density parity-check (LDPC) codes are considered \textit{fixed-rate} channel codes since they incorporate a fixed amount of redundant information \cite{Bonello_11}. However, there exist some well-known techniques for adapting the coding rate of a linear code, one of them is \textit{puncturing}. A linear code is punctured by deleting a set of symbols in a codeword. 

When puncturing a code, we must differentiate between random and intentional puncturing. In the former, punctured symbols are randomly chosen, whereas in the latter, the code is analyzed to select the set of symbols to puncture. The asymptotic performance of random and intentional punctured LDPC codes was analyzed in~\cite{Ha_04, Hsu_08}, and puncturing thresholds were also identified in~\cite{Pishro-Nik_07}. Some other methods delved into the code structure to identify good puncturing patterns \cite{Ha_06, Vellambi_09}, or examined its graph construction \cite{Hu_05, Bandi_11} to facilitate its puncturing \cite{Yazdani_04, Tian_05, Kim_09, Shi_09}.

The objectives pursued when switching from random puncturing can lead to different solutions. In particular, algorithms that focus in covering a wide range of coding rates do not offer the best performance for small puncturing proportions and vice-versa. Several algorithms find puncturing patterns that allow to cover a wide range of rates \cite{Ha_06, Ha_07, Park_07, Vellambi_09, El-Khamy_09}. However, when working with short length codes, but also in other scenarios, the ensemble of punctured symbols determines the decoding performance. In this work, we describe an algorithm that we call \textit{untainted}. 
Its main focus is optimizing the decoding of moderately punctured codes.

The rest of this paper is organized as follows. In Section~\ref{sec:algorithm}, we introduce the notation, some puncturing properties and the untainted algorithm. In Section~\ref{sec:results}, we present simulation results over several channels, and compare them with the results in previous studies \cite{Ha_06, Vellambi_09}. Conclusions are presented in Section~\ref{sec:conclusions}.

%% file: notation.tex
\section{Untainted Puncturing}
\label{sec:algorithm}

\subsection{Notation and Definitions}

Error correcting codes can be represented by bipartite graphs linking symbol nodes with check nodes. Let\footnote{We follow a notation similar to that in Ha \textit{et al.}~\cite{Ha_06}} $\mathcal{N}(z)$ denote the neighborhood of a node $z$, that is, the set of nodes adjacent to $z$. The degree of a node is defined as the cardinality of its neighborhood. This concept can be extended to include all the nodes reachable from $z$ by traversing a maximum of $k$ edges, we call this set of nodes $\mathcal{N}^k(z)$ the neighborhood of depth $k$ of $z$.

Let $\mathcal P$ stand for the set of punctured symbols, $v\in\mathcal P$ belongs to the set of $1$-step extended recoverable symbols ($v\in\mathcal{R}_1$) if $\exists c\in\mathcal{N}(v)$ such that $\forall w\in\mathcal{N}(c)\backslash \{v\}, w\notin\mathcal{P}$. For $n>1$ we recursively define the sets of $k$-step extended recoverable symbols $R_k$. We say that a punctured symbol $v\not\in \mathcal{R}_1\cup...\cup\mathcal{R}_{k-1}$ belongs to the set of $k$-step extended recoverable ($v\in\mathcal{R}_k$) symbols if $\exists c\in\mathcal{N}(v)$ and $\exists w\in\mathcal{N}(c)\backslash\{v\}$ such that $ w\in\mathcal{R}_{k-1}$ and $\forall w'\in\mathcal{N}(c)\backslash\{v,w\}, w'\in\mathcal P\Rightarrow w'\in\mathcal{R}_1\cup...\cup\mathcal{R}_{k-1}$.

The graph subjacent to $\mathcal{N}^{2k}(v)$, $v\in\mathcal{R}_k$, is assumed to be tree-like. Let $u$ be a node in  $\mathcal{N}^{2k}(v)$. We denote by ${\mathcal{N}_\downarrow}^t(u)$  the neighborhood of $u$ of depth $t$ restricted to the descendants of $u$ in this tree. Note that ${\mathcal{N}_\downarrow}(u)={\mathcal{N}_\downarrow}^1(u)$. 
We can prune it by eliminating the connection of any symbol $w\in\mathcal{R}_{l}$ with a check $c\in\mathcal{N}_\downarrow(w)$ if $\max_{w'\in{\mathcal{N}_\downarrow}(c)}\left\{m|w'\in\mathcal{R}_{m}\right\}>l-1$. We call this graph $\mathcal T_v$ the \textit{extended recovery tree} of $v$.

We consider in this letter the sum-product decoding algorithm. The algorithm exchanges messages representing probabilities or the log-likelihood ratio (LLR) of probabilities. The messages are iteratively exchanged from symbol to check nodes and from check to symbol nodes. If the decoding graph is tree-like then, for uniform sources and output symmetric channels, sum-product decoding is equivalent to maximum a posteriori decoding and the decoder minimizes the decoding error~\cite{Richardson_01}.

Let us analyze the effect of puncturing on the messages exchanged in the sum-product algorithm. For every $v\in\mathcal P$ the decoding algorithm has no information on the value that $v$ takes. In consequence, for binary input channels it can take both values (one and zero) with probability one half. On the LLR version of the sum-product algorithm the outgoing messages from symbol $v$ on iteration one are equal to zero. 

A check $c\in\mathcal{N}(v)$ in the neighborhood of a punctured symbol $v$ is called a survived check node if $\exists w\in\mathcal{N}_\downarrow(c)|w\in\mathcal{R}_{k-1}$, and a dead check node otherwise. The message that a dead check node $c$ sends to $v$ is a zero LLR. A punctured symbol is recovered when it receives a message from a survived check node. It follows from the previous definitions that a symbol in $\mathcal R_k$ is recovered after $k$ decoding iterations. 

We define the extended recovery error probability of a punctured symbol $P_e(v)$ as the decoding error probability of $v\in\mathcal{R}_k$ after $k$ iterations on $\mathcal T_v$. Assuming that a codeword is sent through a binary input memoryless output symmetric channel the probability of error is independent of the codeword sent~\cite{Richardson_01}. Then $P_e(v)$ is the probability that $v$ takes the value one conditional to sending the all zero codeword.

These concepts, are extended in the sense that they generalize the definitions introduced by Ha \textit{et al.} in~\cite{Ha_06}. The difference between both sets of definitions is that every punctured symbol is connected to only one survived check node in the (non-extended) recovery tree in~\cite{Ha_06} and, in consequence, the (non-extended) recovery error probability is given by the probability that the message sent by the survived node is wrong.

%% file: properties.tex
\subsection{Properties of the extended recovery error probability}

Ha considered the exact recovery error probability over the (non-extended) recovery tree for the binary erasure channel (BEC), the additive white Gaussian noise (AWGN) channel and the binary symmetric channel (BSC). This error probability is a monotone increasing function on the number of nodes in the recovery tree. The algorithm in~\cite{Ha_06} was developed to exploit this non-intuitive property. 

The single survived check node assumption captures the tree structure when a high proportion of symbols are punctured. However, for a low proportion of punctured symbols there can be more than one survived check node. We now show that having more than one survived check node is a desirable property. More precisely, adding a survived check node in an extended recovery tree can not increase $P_e(v)$. We first prove a stronger claim on the BEC, i.e. adding a survived check node decreases $P_e(v)$. Then, we prove the property for general symmetric channels. The idea behind the general proof is that we can reduce the number of survived check nodes by adding noise to the symbol nodes under a survived check node. Then, given that the sum-product algorithm on tree like graphs with uniform priors is equivalent to a maximum a posteriori estimation, the decoding on the noisier tree can not reduce the decoding error probability. 

\begin{theorem}
\label{th:bec}

Let $l,k\in\mathbb{N}$ and $0\leq l<k$. Now consider the subgraph of a check node $z$ of depth $2k-2l-1$ such that $ \max_{w\in\mathcal{N}(z)}\left\{ m|w\in\mathcal{R}_m\right\} = k-l$. Let $\mathcal{T}_{v_1},\mathcal{T}_{v_2}$ be the extended recovery trees associated with punctured symbols ${v_1,v_2}\in\mathcal{R}_k$. Let both trees be identical except for some $x\in\mathcal{N}^{2l}({v_2})$ that is linked with $z$. Then the recovery error probability of ${v_1}$ and ${v_2}$ sent through a BEC($\alpha$), with $0<\alpha < 1$, verify:

\begin{equation}
P_e({v_1}) > P_e({v_2}).
\end{equation}

\end{theorem}

\begin{proof}
The initial erasure probability of a symbol $v$ is:

\begin{equation}
\epsilon_{v}^{(0)}=\left\{
	\begin{array}{ll}
		1 & \mbox{if } v\in\mathcal{P}\\
		\alpha & \mbox{otherwise}
	\end{array}
\right.\nonumber
\end{equation}

\noindent further, $\epsilon$ can be recursively defined for any symbol and check in the tree from its children erasure probabilities:

\begin{equation}
\epsilon_{v}=\epsilon^{(0)}_{v}\prod_{c\in\mathcal{N}_\downarrow(v)} \epsilon_{c}
\end{equation}

\begin{equation}
\epsilon_{c}=1-\prod_{v\in\mathcal{N}_\downarrow(c)}(1-\epsilon_{v}).
\end{equation}

Now, taking into account that there are no punctured symbols within the leave nodes by the definition of the extended recovery tree, it holds that $\epsilon_{v}^{(0)}<1$ for the leaf symbols of the tree spanning from check $z$. It follows by induction that: 1) $\forall v,c\in{\mathcal{N}_\downarrow}^{2k-2l-1}(z),\epsilon_v,\epsilon_c< 1$; which implies that $\epsilon_z<1$, and 2) if we attach a check node $z$ with $\epsilon_z<1$ to a symbol $x$ then $\epsilon_{v_1}>\epsilon_{v_2}$. Finally, the recovery error probability of a symbol node $v$ is $P_e(v) = \epsilon_v/2$, which completes the proof.
\end{proof}

\begin{theorem}

The recovery error probability of ${v_1}$ and ${v_2}$ over the trees $\mathcal{T}_{v_1},\mathcal{T}_{v_2}$  defined exactly as in Th.~\ref{th:bec} and sent through any binary input symmetric output memoryless channel $C$, verify:

\begin{equation}
P_e({v_1})\geq P_e({v_2}).
\end{equation}

\end{theorem}

\begin{proof}
For a precise characterization of $P_e$ in the general setting, we need to track a message density instead of a scalar. The initial density function of a non-recovered punctured node takes the form of the Dirac delta function $D(y)=\delta(y)$, since a punctured node transmits a zero LLR with probability one. Let the remaining nodes in $\mathcal{T}_{{v_1}},\mathcal{T}_{v_2}$  have initial densities given by $P_0(y)$, the initial LLR density associated with channel $C$. 

Now consider a second scenario for $\mathcal{T}_{v_2}$. We associate every leaf node in $\mathcal{N}_\downarrow^{2k-2l-1}(z)$ with samples from $D(y)$, which is equivalent to puncturing these nodes. If we puncture the leave nodes, their parent check nodes do not become survived check nodes and the symbols $w\in\mathcal{N}_\downarrow^{2k-2l-3}(z)|w\in\mathcal{P}$ are not recovered. It follows by induction that $z$ remains a dead check node, i.e. associating the leaves with $D(y)$ is equivalent to eliminating the edge joining $z$ to $x$. In consequence, the density of messages reaching the root node $v_2$ in the second scenario is identical to the density of messages reaching ${v_1}$ in the first scenario. 

The (binary output) degenerate channel $D$, with initial density $D(y)$, transforms any input into a one or a zero with equal probability, i.e. $p_D(1|x)=p_D(0|x)=0.5$. Let $Q$ represent the concatenated channel of $C$ with $D$:

\begin{eqnarray}
\label{eq:sto}
p_Q(y'|x) &=& \sum_{y\in\mathcal Y}p_D(y'|y)p_C(y|x)\nonumber\\ 
         &=& 0.5 \sum_{y\in\mathcal Y} p_C(y|x) = p_D(y'|x).
\end{eqnarray}

In other words,  $D$ can be regarded as the concatenation of $C$ with itself, and the samples from $D(y)$ are stochastically degraded samples of $P_0(y)$~\cite{Cover_91}.

Following the argument in~\cite[Th. 5]{Richardson_01b}, Eq.~(\ref{eq:sto}) implies that $P_e({v_1})\geq P_e({v_2})$. The assertion follows from the fact that the estimate of both $v_1$ and $v_2$ are maximum likelihood estimates.
\end{proof}

%% file: algorithm.tex
\subsection{Untainted Puncturing Algorithm Description}

We introduce the concept of \textit{untainted}, to propose a simple method that chooses symbols such that all the check nodes of a selected symbol are survived nodes. This restriction guarantees that more than one survived check node is associated with every punctured symbol.

\begin{definition}
A symbol node $v$ is said to be \textit{untainted} if there are no punctured symbols within $\mathcal{N}^{2}(v)$.
\end{definition}

Let $\mathcal{X}_{\infty}$ be the set of untainted symbol nodes. Initially, when there are no punctured symbols, $\mathcal{X}_{\infty}$ consists of every symbol node.

\medskip

\begin{algorithmic}

\STATE \COMMENT{Initialize} $\mathcal{X}_\infty = \{1, ..., n\}$, $p=1$.

\WHILE{$\mathcal{X}_\infty \ne \emptyset$}

\STATE \COMMENT{\textit{Step 1.-- Look for candidates}}

\STATE Make the set of candidates  $\Omega$, which is a subset of $\mathcal{X}_{\infty}$, such that $u \in \Omega$ if $|\mathcal{N}^{2}(u)| \le |\mathcal{N}^{2}(v)|$ for any $v \in \mathcal{X}_{\infty}$.

\STATE \COMMENT{\textit{Step 2.-- Select for puncturing}}

\STATE Pick a symbol node $v^{(p)}$ from $\Omega$ (pick one randomly if there exist more that one symbols in $\Omega$).

\STATE \COMMENT{\textit{Step 3.-- Update the set of untainted symbols}}

\STATE $\mathcal{X}_{\infty} = \mathcal{X}_\infty \backslash \mathcal{N}^{2}(v)$

\STATE $p = p + 1$

\ENDWHILE

\end{algorithmic}

\medskip

The algorithm obtains a set of puncturable symbol nodes 
consisting in the symbols selected in the second step. It concludes when  $\mathcal{X}_\infty = \emptyset$, i.e. there is no untainted symbols. The range of values for $p$, the number of punctured symbols, can be found empirically by simulations (see Table~\ref{tab:p-max}).

Note that for codes with an almost regular check node degree distribution, the searching criterion in Step 1 can be simplified: instead of looking for a symbol with the smallest neighboring set of depth 2, the algorithm can look for symbols with the lowest degree.

%% file: results.tex
\section{Simulation Results}
\label{sec:results}

The untainted algorithm ensures that the extended recovery tree of a punctured symbol has more than one survived check node. In this section, we construct codes to show that the untainted algorithm yields a better performance in terms of the frame error rate (FER). We have constructed $10^{4}$ bit-long irregular LDPC codes of different coding rates for the BEC, the BSC and the AWGN channels. The polynomials for the BSC with rates $(0.5,0.6,0.7,0.8)$ have been drawn from \cite{Martinez_12}. The remaining generating polynomials, as well as all the matrices used can be checked in \cite{gicc}.

\begin{figure}[t]
\includegraphics[width=\linewidth]{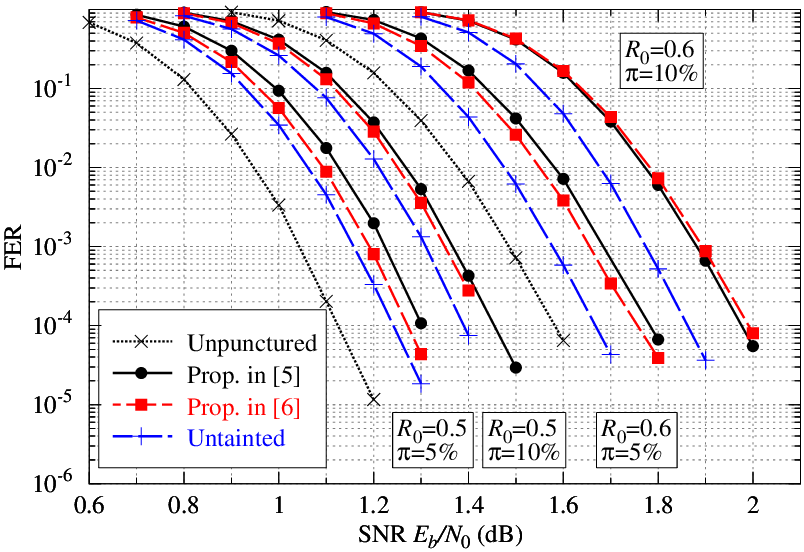}
\caption{FER over the AWGN as a function of the signal-to-noise ratio (SNR). Two LDPC codes with coding rates $R_0 = 0.5$ and $R_0 = 0.6$, and two different proportions of punctured symbols, $\pi = 5\%$ and $\pi = 10\%$ were used.}
\label{fig:fer-awgn}
\end{figure}

\begin{figure}[t]
\includegraphics[width=\linewidth]{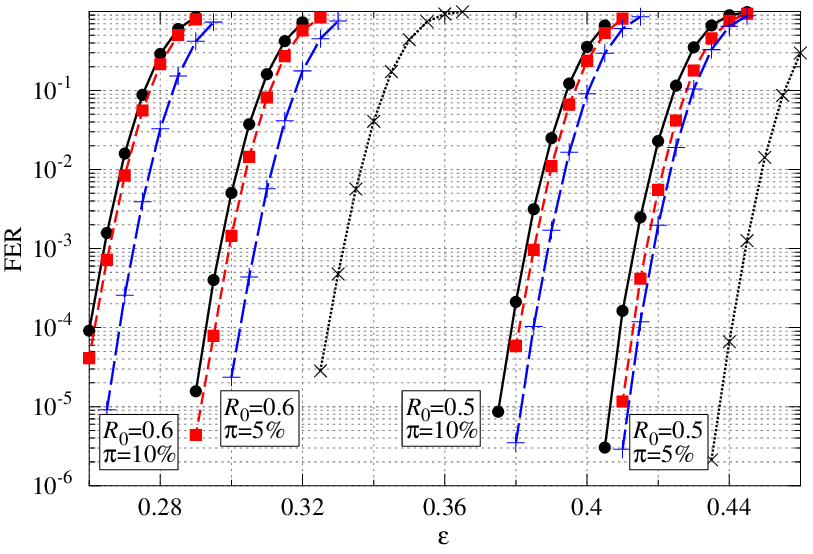}
\caption{FER over the BEC with  crossover probability $\varepsilon$ for different intentional puncturing strategies. Two LDPC codes with coding rates $R_0 = 0.5$ and $R_0 = 0.6$, and two different proportions of punctured symbols, $\pi = 5\%$ and $\pi = 10\%$ were used. See  inset in Fig.~\ref{fig:fer-awgn} for symbols and line-styles.}
\label{fig:fer-bec}
\end{figure}

\begin{figure*}[t]
\includegraphics[width=\linewidth]{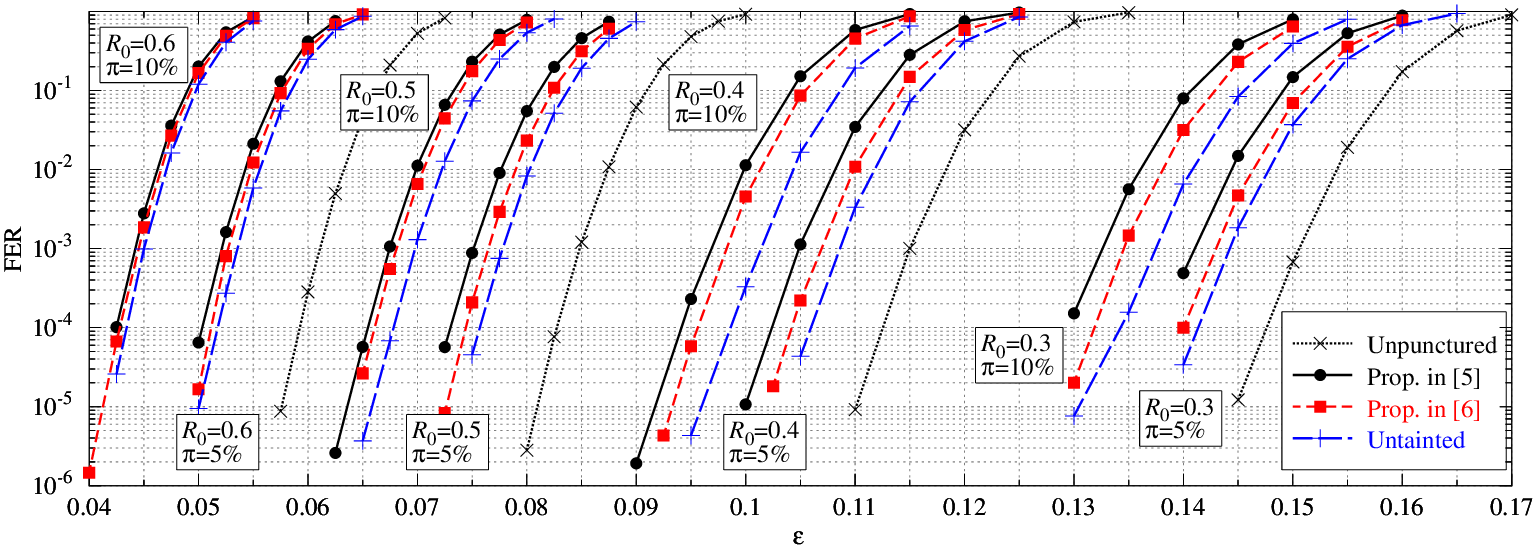}
\caption{FER over the BSC with crossover a probability $\varepsilon$ for several LDPC codes with coding rates $R_0 = 0.3$, $R_0 = 0.4$, $R_0 = 0.5$ and $R_0 = 0.6$, and two different proportions of punctured symbols, $\pi = 5\%$ and $\pi = 10\%$.}
\label{fig:fer-bsc}
\end{figure*}

Figs.~\ref{fig:fer-awgn}, \ref{fig:fer-bec} and \ref{fig:fer-bsc} compare the performance of intentional punctured codes using the untainted algorithm with those in~\cite{Ha_06} and in~\cite{Vellambi_09} for different proportions of punctured symbols $\pi$. All codes are decoded with 200 iterations using the sum-product algorithm~\cite{Richardson_01} over the graph of the mother (non-punctured) code.

As in Ha \textit{et al.}~\cite{Ha_06} we used three random seeds for the simulations of each intentional puncturing algorithm, the results in the figures show the intermediate performer. 

The untainted punctured LDPC codes outperform the codes punctured following the algorithms in~\cite{Ha_06} and in~\cite{Vellambi_09} for all the coding rates, channels and puncturing proportions considered. For a FER of $10^{-3}$ the strongest improvements appear in the BSC for a mother code of rate $R_0=0.3$ punctured a 10\%, in the BEC for a mother code of rate $R_0=0.6$ punctured a 5\% and in the AWGN for a mother code of rate $R_0=0.6$ punctured a 10\% respectively.

Table~\ref{tab:p-max} shows $p_{\mathrm{min}}$ and $p_{\mathrm{max}}$, the minimum and maximum values of $p$, respectively, after $5\cdot 10^3$ algorithm runs. Sizes are computed for both the untainted algorithm and the algorithm in~\cite{Ha_06}. The values are computed for the same codes used in Fig.~\ref{fig:fer-bsc} and two additional codes of rates $0.7$ and $0.8$. The table shows that the untainted algorithm can puncture a smaller number of symbols compared to~\cite{Ha_06}. This behavior is consistent with the additional number of survived check nodes required by the untainted algorithm.

\begin{table}
\caption{Number of Punctured Symbols}
\label{tab:p-max}
\centering
\begin{tabular}{llcccccc}
 & & \multicolumn{6}{c}{Coding rate (mother code)} \\
  & & 0.3 & 0.4 & 0.5 & 0.6 & 0.7 & 0.8 \\
\hline
Ref.~\cite{Ha_06} & $p_\mathrm{min}$ & 4628 & 4031 & 3439 & 2866 & 2258 & 1581 \\
 & $p_\mathrm{max}$ & 4753 & 4139 & 3552 & 2983 & 2353 & 1657 \\
\hline
Untainted & $p_\mathrm{min}$ & 2603 & 2286 & 1909 & 1587 & 1212 & 851 \\
 & $p_\mathrm{max}$ & 2686 & 2374 & 1987 & 1655 & 1273 & 901 \\
\hline
\end{tabular}
\end{table}

%% file: conclusions.tex
\section{Conclusions}
\label{sec:conclusions}

We proved that having more than one survived check node in the extended recovery tree is a desirable property. The untainted algorithm is a method that chooses symbols such that all the check nodes of a selected symbol are survived check nodes. Furthermore, the algorithm can be implemented with a low computational cost. 

Simulation results show that the performance of the untainted algorithm in terms of the FER is better than the best  intentional puncturing algorithms in the literature for a range of coding rates, channels and puncturing proportions. The drawback of using this algorithm is a reduction in the maximum achievable rate since the proportion of puncturable symbols is limited by the untainted criterion.